\documentclass
	[
		10 pt
	]
	{article}
\usepackage
	[
		a4paper,
		left = 2.500 cm,
		right = 2.500 cm,
		top = 2.500 cm,
		bottom = 3.000 cm
	]
	{geometry}
\usepackage{tikz, pgf, pgfplots}
\usetikzlibrary
	{
		3d,
		arrows,											
		arrows.meta,
		automata,										
		backgrounds,
		bending,
		calc,
		chains,
		datavisualization.formats.functions,			
		decorations.pathmorphing,
		decorations.pathreplacing,
		decorations.text,
		fadings,
		fit,
		graphs,
		graphs.standard,
		matrix,
		patterns,
		positioning,									
		quantikz2,
		quotes,
		shadings,
		shadows,
		shadows.blur,
		shapes,
		shapes.geometric,
		trees
	}
\pgfplotsset{compat = newest}
\usepgflibrary
	{
		shapes.misc
	}
\usepgfplotslibrary
	{
		dateplot
	}
\usepackage{tikzpeople}
\usepackage{fontawesome6}
\usepackage{smartdiagram}

\allowdisplaybreaks
\usepackage{amsthm}
\usepackage{amssymb}
\usepackage[bb = boondox]{mathalfa}
\usepackage{dsfont}
\usepackage{newtxmath}
\usepackage{mathtools}
\usepackage{multicol}
\usepackage{physics2}
\usephysicsmodule{ab}									
\usephysicsmodule{braket}								
\usephysicsmodule{diagmat}								
\usephysicsmodule{nabla.legacy}							
\usepackage{empheq}
\usepackage[ compat = newest ]{yquant}

\usepackage{sectsty}
\allsectionsfont{\color{WordBlue}}
\usepackage[x11names]{xcolor}
\usepackage{varwidth}
\usepackage{graphicx}
\usepackage{wrapfig}
\usepackage{subcaption}
\usepackage{xurl}                                                                                  %
\usepackage{hyperref}
\hypersetup
	{
		colorlinks,
		linkcolor = {RedPurple},
		citecolor = {WordBlueDarker25},
		urlcolor = {WordAquaDarker50}
	}
\usepackage{tabularray}
\usepackage{colortbl}
\usepackage{float}
\usepackage{nicematrix}
\usepackage{cellspace}
\usepackage{makecell}
\usepackage{multirow}
\usepackage{caption}
\usepackage[ linesnumbered, tworuled, vlined ] {algorithm2e}
\usepackage{appendix}
\usepackage{enumitem}
\usepackage{siunitx}
\usepackage{xintexpr}
\usepackage{spreadtab}
\usepackage{framed}
\usepackage{svg}
\usepackage{lipsum}
\usepackage{listings}
\definecolor{MyLightRed}{RGB}{244, 213, 245}
\definecolor{Purple}{HTML}{911146}
\definecolor{PurpleDark}{RGB}{102, 0, 102}
\definecolor{RedDarkLight}{HTML}{ea005f}
\definecolor{RedDarkLightest}{HTML}{ff0088}
\definecolor{RedPurple}{HTML}{AA007F}
\definecolor{WordPinkAccent1Darker25}{HTML}{B3186D}
\definecolor{WordPinkAccent1Darker50}{HTML}{781049}
\definecolor{WordPinkAccent1Lighter40}{HTML}{EE80BC}
\definecolor{WordPinkAccent1Lighter60}{HTML}{F3AAD2}
\definecolor{WordPinkAccent1Lighter80}{HTML}{F9D4E8}
\definecolor{WordRed}{RGB}{255, 0, 102}
\definecolor{WordRedAccent5Lighter60}{HTML}{F5B5A7}
\definecolor{WordRedAccent5Darker25}{HTML}{B23214}
\definecolor{GreenDark}{HTML}{225522}
\definecolor{GreenLighter1}{HTML}{00B383}
\definecolor{GreenLighter2}{HTML}{00AA7F}
\definecolor{GreenLightest}{HTML}{00FFA0}
\definecolor{GreenTeal}{HTML}{008080}
\definecolor{WordLightGreen}{RGB}{140, 214, 192}
\definecolor{WordGreen}{RGB}{0, 176, 80}
\definecolor{BlueVeryDark}{HTML}{222255}
\definecolor{MyBlue}{RGB}{0, 64, 128}
\definecolor{MyDarkBlue}{RGB}{0, 51, 102}
\definecolor{MyVeryLightBlue}{RGB}{211, 245, 247}
\definecolor{WordBlue}{RGB}{19, 65, 99}
\definecolor{WordBlueDark}{RGB}{46, 116, 181}
\definecolor{WordBlueDarker}{RGB}{31, 78, 121}
\definecolor{WordBlueDarker25}{RGB}{54, 96, 146}
\definecolor{WordBlueDarker50}{RGB}{36, 64, 98}
\definecolor{WordBlueDarkest}{RGB}{0, 32, 96}
\definecolor{WordBlueLight}{RGB}{0, 112, 192}
\definecolor{WordBlueVeryLight}{HTML}{00B0F0}
\definecolor{WordIceBlue}{RGB}{223, 227, 229}
\definecolor{MagentaDark}{RGB}{106, 65, 152}
\definecolor{MagentaLight}{RGB}{128, 100, 162}
\definecolor{MagentaLighter}{RGB}{161, 106, 221}
\definecolor{MagentaVeryDark}{RGB}{97, 75, 128}
\definecolor{MagentaVeryLight}{RGB}{178, 162, 201}
\definecolor{WordAquaAccent1Darker25}{HTML}{276E8B}
\definecolor{WordAquaAccent1Darker50}{HTML}{1A495D}
\definecolor{WordAquaAccent1Lighter40}{HTML}{7FC0DB}
\definecolor{WordAquaAccent1Lighter60}{HTML}{A9D5E7}
\definecolor{WordAquaAccent1Lighter80}{HTML}{D4EAF3}
\definecolor{WordAquaAccent2Darker25}{HTML}{398E98}
\definecolor{WordAquaAccent2Darker50}{HTML}{265F65}
\definecolor{WordAquaAccent2Lighter40}{HTML}{9AD3D9}
\definecolor{WordAquaAccent2Lighter60}{HTML}{BCE1E5}
\definecolor{WordAquaAccent2Lighter80}{HTML}{DDF0F2}
\definecolor{WordAquaDarker25}{HTML}{31869B}
\definecolor{WordAquaDarker50}{HTML}{215967}
\definecolor{WordAquaLighter40}{HTML}{92CDDC}
\definecolor{WordAquaLighter60}{HTML}{B7DEE8}
\definecolor{WordAquaLighter80}{HTML}{DAEEF3}
\definecolor{WordDarkerTeal}{RGB}{48, 82, 80}
\definecolor{WordDarkTeal}{RGB}{72, 123, 119}
\definecolor{WordDarkTealLighter80}{RGB}{207, 223, 234}
\definecolor{WordLightTeal}{RGB}{160, 199, 197}
\definecolor{WordVeryLightTeal}{RGB}{223, 236, 235}
\definecolor{WordTurquoiseLighter80}{RGB}{209, 238, 249}
\definecolor{Brown}{HTML}{666633}
\definecolor{WordGoldAccent1Darker25}{HTML}{C49A00}
\definecolor{WordGoldAccent1Lighter40}{HTML}{FFDF6A}
\definecolor{WordOrangeAccent2Lighter60}{HTML}{FCD3A4}
\definecolor{WordOrangeAccent4Lighter60}{HTML}{F7C5A1}
\definecolor{LavenderBlush}{RGB}{255, 240, 245}
\definecolor{MediumTurquoise}{RGB}{72, 209, 204}
\definecolor{PowderBlue}{RGB}{176, 224, 230}
\definecolor{SkyBlue}{RGB}{135, 206, 235}
\definecolor{Azure2}{RGB}{224, 238, 238}
\definecolor{Azure3}{RGB}{193, 205, 205}
\definecolor{CadetBlue4}{RGB}{83, 134, 139}
\definecolor{DarkSeaGreen1}{RGB}{193, 255, 193}
\definecolor{DeepPink4}{RGB}{139, 10, 80}
\definecolor{Honeydew2}{RGB}{224, 238, 224}
\definecolor{LightSkyBlue1}{RGB}{176, 226, 255}
\definecolor{LightSkyBlue3}{RGB}{141, 182, 205}
\definecolor{LightSkyBlue4}{RGB}{96, 123, 139}
\definecolor{LightSteelBlue1}{RGB}{202, 225, 255}
\definecolor{LightSteelBlue4}{RGB}{110, 123, 139}
\definecolor{MediumPurple1}{RGB}{171, 130, 255}
\definecolor{PaleTurquoise3}{RGB}{150, 205, 205}
\definecolor{PaleVioletRed3}{RGB}{205, 104, 137}
\definecolor{Purple1}{RGB}{155, 48, 255}
\definecolor{SeaGreen1}{RGB}{84, 255, 159}
\definecolor{SeaGreen2}{RGB}{78, 238, 148}
\definecolor{SeaGreen3}{RGB}{67, 205, 128}
\definecolor{SkyBlue1}{HTML}{87CEFF}
\definecolor{SkyBlue4}{RGB}{74, 112, 139}
\definecolor{SteelBlue1}{RGB}{99, 184, 255}
\definecolor{Thistle3}{RGB}{205, 181, 205}
\definecolor{Turquoise4}{RGB}{0, 134, 139}
\definecolor{VioletRed1}{RGB}{255, 62, 150}
\definecolor{VioletRed2}{RGB}{208, 32, 144}
\definecolor{VioletRed3}{RGB}{199, 21, 133}
\definecolor{VioletRed4}{RGB}{139, 10, 80}
\usepackage
	[
		most
	]
	{tcolorbox}
\newcounter{MyCorollary}[section]
\renewcommand{\theMyCorollary}{\thesection.\arabic{MyCorollary}}
\newtcolorbox{corollary} [ 1 ] [ ]
	{
		breakable,
		enhanced,
		enhanced jigsaw,
		skin = enhanced,
		attach boxed title to top left = { xshift = -5.000 mm, yshift = 0.000 mm },
		boxed title style = { boxrule = 0.000 pt, sharp corners = all },
		colbacktitle = RedPurple!90!black,
		coltitle = white,
		fonttitle = \bfseries,
		varwidth boxed title,
		colback = RedPurple!03,
		colframe = RedPurple,
		sharp corners = all,
		toprule = 0.000 mm,
		bottomrule = 0.500 mm,
		leftrule = 0.500 mm,
		rightrule = 0.000 mm,
		code = { \refstepcounter{MyCorollary} },
		title = {Corollary~\theMyCorollary:\if\relax\detokenize{#1}\relax\else~#1\fi},
	}
\newtcolorbox{corollary*} [ 1 ] [ ]
	{
		breakable,
		enhanced,
		enhanced jigsaw,
		skin = enhanced,
		attach boxed title to top left = { xshift = -5.000 mm, yshift = 0.000 mm },
		boxed title style = { boxrule = 0.000 pt, sharp corners = all },
		colbacktitle = RedPurple!90!black,
		coltitle = white,
		fonttitle = \bfseries,
		varwidth boxed title,
		colback = RedPurple!03,
		colframe = RedPurple,
		sharp corners = all,
		toprule = 0.000 mm,
		bottomrule = 0.500 mm,
		leftrule = 0.500 mm,
		rightrule = 0.000 mm,
		title = {Corollary\if\relax\detokenize{#1}\relax\else~#1\fi},
	}
\newcounter{MyDefinition}[section]
\renewcommand{\theMyDefinition}{\thesection.\arabic{MyDefinition}}
\newtcolorbox{definition} [ 1 ] [ ]
	{
		breakable,
		enhanced,
		enhanced jigsaw,
		skin = enhanced,
		attach boxed title to top left = { xshift = -5.000 mm, yshift = 0.000 mm },
		boxed title style = { boxrule = 0.000 pt, sharp corners = all },
		colbacktitle = SkyBlue!70,
		coltitle = SkyBlue!30!black,
		fonttitle = \bfseries,
		varwidth boxed title,
		colback = SkyBlue!15,
		colframe = SkyBlue,
		sharp corners = all,
		toprule = 0.000 mm,
		bottomrule = 0.500 mm,
		leftrule = 0.500 mm,
		rightrule = 0.000 mm,
		code = { \refstepcounter{MyDefinition} },
		title = {Definition~\theMyDefinition:\if\relax\detokenize{#1}\relax\else~#1\fi},
	}
\newtcolorbox{definition*} [ 1 ] [ ]
	{
		breakable,
		enhanced,
		enhanced jigsaw,
		skin = enhanced,
		attach boxed title to top left = { xshift = -5.000 mm, yshift = 0.000 mm },
		boxed title style = { boxrule = 0.000 pt, sharp corners = all },
		colbacktitle = SkyBlue!70,
		coltitle = SkyBlue!30!black,
		fonttitle = \bfseries,
		varwidth boxed title,
		colback = SkyBlue!15,
		colframe = SkyBlue,
		sharp corners = all,
		toprule = 0.000 mm,
		bottomrule = 0.500 mm,
		leftrule = 0.500 mm,
		rightrule = 0.000 mm,
		title = {Definition\if\relax\detokenize{#1}\relax\else~#1\fi},
	}
\newcounter{MyExample}[section]
\renewcommand{\theMyExample}{\thesection.\arabic{MyExample}}
\newtcolorbox{example} [ 1 ] [ ]
	{
		breakable,
		enhanced,
		enhanced jigsaw,
		skin = enhanced,
		attach boxed title to top left = { xshift = -5.000 mm, yshift = 0.000 mm },
		boxed title style = { boxrule = 0.000 pt, sharp corners = all },
		colbacktitle = WordAquaAccent1Darker25,
		coltitle = white,
		fonttitle = \bfseries,
		varwidth boxed title,
		colback = WordAquaAccent1Lighter80!25,
		colframe = WordAquaAccent1Darker25,
		sharp corners = all,
		toprule = 0.000 mm,
		bottomrule = 0.500 mm,
		leftrule = 0.500 mm,
		rightrule = 0.000 mm,
		code = { \refstepcounter{MyExample} },
		title = {Example~\theMyExample:\if\relax\detokenize{#1}\relax\else~#1\fi},
	}
\newtcolorbox{example*} [ 1 ] [ ]
	{
		breakable,
		enhanced,
		enhanced jigsaw,
		skin = enhanced,
		attach boxed title to top left = { xshift = -5.000 mm, yshift = 0.000 mm },
		boxed title style = { boxrule = 0.000 pt, sharp corners = all },
		colbacktitle = WordAquaAccent1Darker25,
		coltitle = white,
		fonttitle = \bfseries,
		varwidth boxed title,
		colback = WordAquaAccent1Lighter80!25,
		colframe = WordAquaAccent1Darker25,
		sharp corners = all,
		toprule = 0.000 mm,
		bottomrule = 0.500 mm,
		leftrule = 0.500 mm,
		rightrule = 0.000 mm,
		title = {Example\if\relax\detokenize{#1}\relax\else~#1\fi},
	}
\newcounter{MyLemma}[section]
\renewcommand{\theMyLemma}{\thesection.\arabic{MyLemma}}
\newtcolorbox{lemma} [ 1 ] [ ]
	{
		breakable,
		enhanced,
		enhanced jigsaw,
		skin = enhanced,
		attach boxed title to top left = { xshift = -5.000 mm, yshift = 0.000 mm },
		boxed title style = { boxrule = 0.000 pt, sharp corners = all },
		colbacktitle = PaleVioletRed3!50,
		coltitle = black,
		fonttitle = \bfseries,
		varwidth boxed title,
		colback = WordPinkAccent1Lighter80!12,
		colframe = WordPinkAccent1Darker50,
		sharp corners = all,
		toprule = 0.000 mm,
		bottomrule = 0.500 mm,
		leftrule = 0.500 mm,
		rightrule = 0.000 mm,
		code = { \refstepcounter{MyLemma} },
		title = {Lemma~\the\theMyLemma:\if\relax\detokenize{#1}\relax\else~#1\fi},
	}
\newtcolorbox{lemma*} [ 1 ] [ ]
	{
		breakable,
		enhanced,
		enhanced jigsaw,
		skin = enhanced,
		attach boxed title to top left = { xshift = -5.000 mm, yshift = 0.000 mm },
		boxed title style = { boxrule = 0.000 pt, sharp corners = all },
		colbacktitle = PaleVioletRed3!50,
		coltitle = black,
		fonttitle = \bfseries,
		varwidth boxed title,
		colback = WordPinkAccent1Lighter80!12,
		colframe = WordPinkAccent1Darker50,
		sharp corners = all,
		toprule = 0.000 mm,
		bottomrule = 0.500 mm,
		leftrule = 0.500 mm,
		rightrule = 0.000 mm,
		title = {Lemma\if\relax\detokenize{#1}\relax\else~#1\fi},
	}
\newcounter{MyProposition}[section]
\renewcommand{\theMyProposition}{\thesection.\arabic{MyProposition}}
\newtcolorbox{proposition} [ 1 ] [ ]
	{
		breakable,
		enhanced,
		enhanced jigsaw,
		skin = enhanced,
		attach boxed title to top left = { xshift = -5.000 mm, yshift = 0.000 mm },
		boxed title style = { boxrule = 0.000 pt, sharp corners = all },
		colbacktitle = cyan7!50,
		coltitle = black,
		fonttitle = \bfseries,
		varwidth boxed title,
		colback = cyan9!25,
		colframe = cyan5,
		sharp corners = all,
		toprule = 0.000 mm,
		bottomrule = 0.500 mm,
		leftrule = 0.500 mm,
		rightrule = 0.000 mm,
		code = { \refstepcounter{MyProposition} },
		title = {Proposition~\theMyProposition:\if\relax\detokenize{#1}\relax\else~#1\fi},
	}
\newtcolorbox{proposition*} [ 1 ] [ ]
	{
		breakable,
		enhanced,
		enhanced jigsaw,
		skin = enhanced,
		attach boxed title to top left = { xshift = -5.000 mm, yshift = 0.000 mm },
		boxed title style = { boxrule = 0.000 pt, sharp corners = all },
		colbacktitle = cyan7!50,
		coltitle = black,
		fonttitle = \bfseries,
		varwidth boxed title,
		colback = cyan9!25,
		colframe = cyan5,
		sharp corners = all,
		toprule = 0.000 mm,
		bottomrule = 0.500 mm,
		leftrule = 0.500 mm,
		rightrule = 0.000 mm,
		title = {Proposition\if\relax\detokenize{#1}\relax\else~#1\fi},
	}
\newcounter{MyTheorem}[section]
\renewcommand{\theMyTheorem}{\thesection.\arabic{MyTheorem}}
\newtcolorbox{theorem} [ 1 ] [ ]
	{
		breakable,
		enhanced,
		enhanced jigsaw,
		skin = enhanced,
		attach boxed title to top left = { xshift = -5.000 mm, yshift = 0.000 mm },
		boxed title style = { boxrule = 0.000 pt, sharp corners = all },
		colbacktitle = WordAquaAccent2Darker25,
		coltitle = white,
		fonttitle = \bfseries,
		varwidth boxed title,
		colback = WordAquaAccent2Lighter80,
		colframe = WordAquaAccent2Darker25,
		sharp corners = all,
		toprule = 0.000 mm,
		bottomrule = 0.500 mm,
		leftrule = 0.500 mm,
		rightrule = 0.000 mm,
		code = { \refstepcounter{MyTheorem} },
		title = {Theorem~\theMyTheorem:\if\relax\detokenize{#1}\relax\else~#1\fi},
	}
\newtcolorbox{theorem*} [ 1 ] [ ]
	{
		breakable,
		enhanced,
		enhanced jigsaw,
		skin = enhanced,
		attach boxed title to top left = { xshift = -5.000 mm, yshift = 0.000 mm },
		boxed title style = { boxrule = 0.000 pt, sharp corners = all },
		colbacktitle = WordAquaAccent2Darker25,
		coltitle = white,
		fonttitle = \bfseries,
		varwidth boxed title,
		colback = WordAquaAccent2Lighter80,
		colframe = WordAquaAccent2Darker25,
		sharp corners = all,
		toprule = 0.000 mm,
		bottomrule = 0.500 mm,
		leftrule = 0.500 mm,
		rightrule = 0.000 mm,
		title = {Theorem\if\relax\detokenize{#1}\relax\else~#1\fi},
	}
\newcounter{mathseed}
\pgfmathsetseed{\arabic{mathseed}}
\pgfdeclarelayer{background}
\pgfsetlayers{background, main}
\pgfdeclaredecoration{irregular fractal line}{init}
{
	\state{init}[width=\pgfdecoratedinputsegmentremainingdistance]
	{
		\pgfpathlineto{\pgfpoint{random*\pgfdecoratedinputsegmentremainingdistance}{(random*\pgfdecorationsegmentamplitude-0.02)*\pgfdecoratedinputsegmentremainingdistance}}
		\pgfpathlineto{\pgfpoint{\pgfdecoratedinputsegmentremainingdistance}{0pt}}
	}
}
\def\tornpaper#1{%
	\ifthenelse{\isodd{\value{mathseed}}}
	{%
		\tikz
		{
			\node[inner sep = 1em] (A) {#1};		
			\begin{pgfonlayer}{background}			
				\fill[paper]						
				\pgfextra{\pgfmathsetseed{\arabic{mathseed}}\addtocounter{mathseed}{1}}%
				{decorate[irregular cloudy border]{decorate{decorate{decorate{decorate[ragged border]{
										(A.north west) -- (A.north east)
				}}}}}}
				-- (A.south east)
				\pgfextra{\pgfmathsetseed{\arabic{mathseed}}}%
				{decorate[irregular spiky border]{decorate{decorate{decorate{decorate[ragged border]{
										-- (A.south west)
				}}}}}}
				-- (A.north west);
			\end{pgfonlayer}
		}
	}
	{%
		\tikz{
			\node[inner sep=1em] (A) {#1};  
			\begin{pgfonlayer}{background}  
				\fill[paper] 
				\pgfextra{\pgfmathsetseed{\arabic{mathseed}}\addtocounter{mathseed}{1}}%
				{decorate[irregular spiky border]{decorate{decorate{decorate{decorate[ragged border]{
										(A.north east) -- (A.north west)
				}}}}}}
				-- (A.south west)
				\pgfextra{\pgfmathsetseed{\arabic{mathseed}}}%
				{decorate[irregular cloudy border]{decorate{decorate{decorate{decorate[ragged border]{
										-- (A.south east)
				}}}}}}
				-- (A.north east);
		\end{pgfonlayer}}
	}
}
\usepackage [ scale = 2 ] {ccicons}
\usepackage{changepage}
\usepackage{standalone}                                                                            %
\title
	{
		Extending the El Farol Bar Game with Partial Observability and Incentive Design
	}

\usepackage{orcidlink}
\newcommand{\orcidicon}[1]{\href{https://orcid.org/#1}{\includegraphics[height=\fontcharht\font`\B]{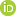}}}

\author
	{
		Iosif Polenakis$^1$\orcidicon{0000-0002-6427-5519},
		Kalliopi Kastampolidou$^2$\orcidicon{0000-0003-3607-9569}
		and
		Theodore Andronikos$^1$\orcidicon{0000-0002-3741-1271}
		\\[10 pt]
		$^1$
		Department of Informatics, Ionian University, \\
		7 Tsirigoti Square, 49100 Corfu, Greece; \\
		\{ ipolenakis,andronikos \}@ionio.gr
		\\[8 pt]
		$^2$
		Institute of Applied Biosciences, \\
		Centre for Research and Technology Hellas (CERTH), Thessaloniki, Greece; \\
		kkastampolidou@certh.gr
	}
\begin{document}

\maketitle

\begin{abstract}
	The El Farol Bar game is a classic model of coordination under uncertainty, traditionally treating the venue as a passive constraint. In this work, we re-conceptualize the problem by modeling the bar as a strategic player equipped with AI-driven learning capabilities. We extend the original framework to include partial observability, i.e., agents observe only subsets of past attendees, and transform the bar from a passive capacity threshold into an active mechanism designer that adjusts pricing policies to balance revenue, utilization, and sustainability constraints. Agents employ AI-based learning to form beliefs and adapt attendance strategies under incomplete information, while the bar uses policy learning to optimize dynamic pricing. The resulting two-sided learning system frames coordination as a co-evolutionary process between boundedly rational agents and an adaptive institution, offering insights into congestion management, resource allocation, and mechanism design in complex adaptive systems.
\end{abstract}
\section{Introduction} \label{sec: Introduction}

The El Farol Bar problem is a well-known model of coordination under uncertainty, where agents repeatedly decide whether to attend a crowded venue based on limited information about others’ behavior. Traditionally, the bar is treated as a passive entity with a fixed capacity threshold, while agents independently adapt their strategies over time. In this work, we extend the classical framework by introducing partial observability and by modeling the bar itself as an active AI-driven decision maker. Agents employ learning mechanisms to adapt their attendance strategies under incomplete information, while the bar dynamically adjusts pricing policies to optimize objectives such as revenue, utilization, and sustainability. This creates a two-sided adaptive system in which both agents and institution co-evolve through learning and interaction.

\subsection{Congestion Problems Meet Mechanism Design}

Arthur's El Farol Bar game \cite{arthur1994inductive} captures something fundamental about coordination under uncertainty. Every week, people decide whether to visit a bar that is enjoyable only when it's not too crowded. No one can coordinate in advance or know others' plans; they can only look at past attendance and guess. If everyone uses the same logic, the system breaks: either everyone shows up (overcrowding) or no one does (wasted capacity). The problem has become a classic precisely because there's no purely deductive solution.

What made Arthur's formulation influential was the realization that this problem can be considered a problem of learning rather than optimization. Agents do not compute equilibria; they build prediction models from experience and adapt as patterns shift. This framing resonated deeply with AI researchers working on multi-agent systems, where learning algorithms must interact without full information about other agents' strategies or even the environment's structure.

This framework has since appeared across many domains such as traffic routing, bandwidth allocation, load balancing, even high-frequency trading\cite{marsili2000exact,bertolotti2025epidemiological}, while it is also utilized to study how algorithms co-adapt when competing for shared resources and whether their collective behavior approximates an equilibrium.

In our approach, the model could be more realistic if endowed with certain additional important traits. In reality, bars are not passive: they adjust prices, run happy hour specials, impose cover charges when busy or use reservation systems. Their functional costs thus set a minimum revenue thresholds below which they shut down. The classical El Farol game treats the bar as a fixed threshold, ignoring that venues are strategic actors facing sustainability constraints and actively managing demand.

In this work, we study these settings from the bar's perspective. What happens when the venue can respond? When both customers and the institution learn and adapt to each other? When we trade perfect rationality for AI-based learning on both sides? These questions sit at the intersection of game theory, multi-agent learning, and mechanism design, and they matter for understanding everything from dynamic pricing in hospitality to cloud resource allocation.

Games can offer fresh perspectives and additional insights, even in the context of serious and difficult problems. It is worth noting that the applications of games have long exceeded conventional settings and have ventured into the realm of unconventional computing. This is justified by their power in providing a radically novel approach, often leading to unconventional strategies superior to the classical ones. For instance, quantum games have addressed serious problems like cryptographic protocols \cite{Bennett1984, Ampatzis2021, Andronikos2025c}. Likewise bio-inspired games employ biostrategies that perform better than conventional strategies—even in the well-known Prisoners' Dilemma game (see for example \cite{Kastampolidou2020, Kastampolidou2020a, Kastampolidou2021, Kostadimas2021, Papalitsas2021, Kastampolidou2023}). 
This perspective reinforces our belief that the game-theoretic framework may facilitate a deeper understanding of the extended El Farol Bar problem.

\subsection{The Bar as Strategic Player}

This work extends El Farol in three ways. First, each attending agent only possesses partial information, i.e., they observe the other agents present there, but not everyone. This creates asymmetric knowledge across the population. Second, agents do not follow fixed prediction rules; they use AI-based learning to form and update beliefs about attendance patterns. Third, and most importantly, the bar becomes a player in its own right.

The bar faces real constraints. It needs minimum revenue to stay open-fall below that for too long and it goes bankrupt. It wants decent utilization without exceeding capacity or having so few customers that the atmosphere dies. To influence these outcomes, the bar controls pricing: offering discounts during slow periods or charging more when it expects crowds.

This completely transforms the game. It is no longer just customers coordinating with each other. Now you have a two-sided system where customers learn attendance patterns and respond to prices, while the bar learns customer behavior and adjusts prices accordingly. Neither side has perfect information or unlimited computing power. Both use AI to navigate uncertainty and adapt over time.

The structure becomes hierarchical. Customers decide whether to attend given the bar's prices and their beliefs about who else might show up. The bar sets prices anticipating how customers will respond through their learning dynamics. But since customer learning changes the response function itself, the bar is aiming at a moving target.

This connects to broader questions in AI and economics. Can adaptive institutions use mechanism design, i.e., dynamic pricing in this case, to steer decentralized systems toward desirable outcomes? The bar acts as a mechanism designer, using the price as a control variable to shape incentives. Customers respond boundedly rationally, and their collective behavior feeds back into the bar's policy updates. The whole system co-evolves.

\subsection{Road-map and Contribution}

Section 2 reviews the evolution of these models-from Arthur's original bar problem through the minority game's abstraction to the Kolkata restaurant problem with multiple venues. The common thread: all treat venues as passive. We argue that this is a critical limitation.

Section 3 formalizes what it means for the bar to be strategic. We define its strategy space (pricing, capacity, information disclosure), its objectives (revenue, utilization, sustainability), and the constraints it faces (capacity limits, minimum occupancy, bankruptcy threshold). The mathematical model treats this as a two-sided dynamic game.

Section 4 is where AI enters. We describe how customers use AI to forecast demand and form beliefs when they can only observe partial information. They might use Bayesian updating, reinforcement learning, or no-regret algorithms. Then we flip to the bar's side: it uses AI to learn customer response functions, optimize dynamic pricing, and adapt to non-stationary behavior. The section ends by analyzing what happens when both sides learn simultaneously, which leads to co-evolutionary dynamics and feedback loops.

Section 5 tackles equilibrium. Standard Nash equilibrium does not fit well here because the environment is non-stationary and agents use learning rather than perfect optimization. We discuss approximate Bayesian Nash equilibrium, learning-consistent concepts like coarse correlated equilibrium, and what ``AI-augmented equilibrium'' means. We also compare welfare when the bar is strategic versus passive-does institutional strategy improve efficiency or just extract more surplus?

Section 6 wraps up with applications (dynamic pricing in hospitality, cloud computing, ride-sharing) and open questions for future work. \\

The contributions of the proposed approach are listed below:

\begin{itemize}
	\item	From a theoretical viewpoint, we generalize El Farol-type games to incorporate mechanism design and two-sided learning.
	\item	Methodologically, we show how AI learning dynamics can approximate equilibrium behavior without assuming perfect rationality.
	\item	Finally, in terms of practical value, we provide a framework for analyzing sustainability and efficiency in congestion-prone systems where institutions adapt strategically.
\end{itemize}

\section{From Passive Venues to Strategic Bars}

In this section, we discuss the setting of the initial El Farol bar game, the abstraction of the minority game, the Kolkata paise restaurant problem, and the limitations set by the assumption of passive venues.

\subsection{El Farol bar: The Classical Setup}

Arthur's formulation \cite{arthur1994inductive} centers on a bar in Santa Fe with limited capacity. Each week, $N$ potential clients (players) must independently decide whether to attend. The bar is enjoyable if attendance stays below some threshold $L < N$, but becomes unpleasant when overcrowded. Crucially, there is no mechanism for advance coordination or communication, i.e., each person must predict attendance based solely on past observations.

Let $n(t)$ denote the number of attendees in week $t$. Each agent $i$ decides whether to attend, $d_i(t) \in \{0,1\}$, such that:

\begin{equation}
	n(t) = \sum_{i=1}^{N} d_i(t)
\end{equation}

\noindent Agent $i$ receives utility:
%

\begin{equation}
	u_i(t)=
	\begin{cases}
		U, & \text{if } d_i(t)=1 \text{ and } n(t)\leq L,\\
		-C_A, & \text{if } d_i(t)=1 \text{ and } n(t)>L,\\
		0, & \text{if } d_i(t)=0.
	\end{cases}
\end{equation}

\noindent where $U > 0$ represents the enjoyment from attending an uncrowded bar, and $C_A > 0$ represents the cost of attending when overcrowded. The payoff structure creates a coordination problem: attending is only worthwhile if sufficiently few others attend.

Arthur's agents employ inductive reasoning through prediction models. Each agent maintains multiple predictors, i.e., simple rules that forecast next week's attendance based on recent history. Predictors might include ``next week's attendance will equal last week's attendance'' or ``attendance oscillates, so predict the opposite of last week''. Each agent tracks predictor accuracy and uses whichever has performed best recently.

The time-averaged attendance tends to hover near the threshold $L$, with significant fluctuations \cite{arthur1994inductive,bell2002keeping}. When attendance falls below $L$, more agents' predictors suggest attending, pushing attendance upward. When it exceeds $L$, predictors adjust and attendance drops. This creates a self-correcting but never-stable equilibrium. Crucially, predictor diversity prevents the system from locking into inefficient periodic cycles.

\subsection{Minority Game: Abstraction and Symmetry}

The minority game, introduced by Challet and Zhang \cite{challet1997emergence}, abstracts resource competition into its purest form. An odd number of agents must independently choose between two options in each round. Those selecting the minority option win; those in the majority lose. This captures an essential feature of many scenarios: success comes from avoiding overcrowding rather than following the crowd.

Consider $N$ agents (where $N$ is odd) choosing $a_i(t) \in \{-1, +1\}$ at time $t$. The aggregate is:

\begin{equation}
	A(t) = \sum_{i=1}^{N} a_i(t)
\end{equation}

Agents in the minority win. Each agent's payoff is:

\begin{equation}
	u_i(t) = -a_i(t) \cdot \text{sgn}(A(t))
\end{equation}

The game incorporates bounded rationality deliberately. Agents cannot compute optimal strategies. Instead, each maintains a small set of decision rules (strategies) mapping recent historical outcomes to actions \cite{challet1999minority}. The history is encoded as a binary string of length $M$, creating $P = 2^M$ possible ``market states".

Each strategy is a lookup table: for each possible history state $\mu$, the strategy specifies an action. Agents evaluate strategies based on virtual performance-tracking how each strategy would have performed even when not selected. At each time step, the agent uses their highest-scoring strategy. If all agents used identical prediction strategies, the system would either lock into periodic cycles or fail to utilize capacity efficiently. The heterogeneity of prediction methods, i.e., some agents using moving averages, others using pattern recognition, still others using contrarian logic, creates the variability that allows the system to explore different attendance levels \cite{whitehead2008farol,franke2003reinforcement,rand2007farol}.

The system's behavior depends critically on $\alpha = P/N = 2^M/N$-the ratio of available information to competing agents \cite{challet2000statistical}. With few agents ($\alpha$ large), the game exhibits persistent patterns and high volatility. As agents increase (decreasing $\alpha$), competition causes patterns to disappear, volatility drops, and outcomes become unpredictable. This phase transition produces market-like properties-volatility clustering and fat-tailed distributions-without explicit design \cite{johnson1998self, savit1999}.

Minority games have been applied to various domains including market microstructure, traffic routing, and resource allocation \cite{lo2004minority}. However, the standard formulation focuses exclusively on agent behavior and strategy evolution. The question of how the resource provider, i.e., the entity managing the limited resource itself, might respond strategically has received less attention. This perspective becomes particularly relevant when we consider the El Farol Bar problem.

\subsection{Kolkata Paise Restaurant Problem}

The Kolkata Paise Restaurant (KPR) problem, introduced by Chakrabarti et al., \cite{chakrabarti2009kolkata,ghosh2013kolkata}, can be viewed as an extension of the El Farol game from a single bar to multiple restaurants competing for customers. The problem takes its name from small, inexpensive restaurants in Kolkata where lunch crowds create coordination challenges across the city rather than at a single location. Consider $N$ agents and $N$ restaurants, each with capacity for exactly one customer. In each round, every agent chooses one restaurant. If exactly one customer arrives, they are served and receive utility $U > 0$. If multiple customers arrive at the same restaurant, all receive zero.

Let $r_i(t) \in \{1, 2, \ldots, N\}$ denote agent $i$'s restaurant choice. Restaurant $j$'s occupancy is:

\begin{equation}
	n_j(t) = \sum_{i=1}^{N} \mathbb{I}\bigl(r_i(t)=j\bigr)
\end{equation}

Agent $i$'s payoff is:

\begin{equation}
	u_i(t) =
	\begin{cases}
		U, & \text{if } n_{r_i(t)}(t)=1,\\
		0, & \text{otherwise}.
	\end{cases}
\end{equation}

\noindent The social optimum occurs when all agents successfully coordinate to visit different restaurants, achieving utilization $f = 1.0$. Without communication, this is highly unlikely.

\noindent The key metric is utilization-the fraction of customers successfully served:
\begin{equation}
	f(t) = \frac{1}{N} \sum_{j=1}^{N} \mathbb{I}(n_j(t) = 1)
\end{equation}

\noindent Under random selection, expected utilization approaches $1/e \approx 0.368$ as $N \to \infty$-nearly two-thirds waste. With learning, agents can reach $f \approx 0.5$ to $0.8$ \cite{ghosh2010phase,chakraborti2017kolkata,ghosh2017emergence,biswas2024achieving}, but perfect coordination remains elusive.

In closing this subsection, let us also point out that in the numerous treatises of the literature for the KPR problem one can find connections to iconic optimization problems, like the TSP \cite{Kastampolidou2022}, where a distributed scenario that leads to superior utilization is studied, or to totally unconventional implementations, like the one outlined in \cite{ramzan2013three}.

\subsection{Limitation: All Assume Passive Venues}

Despite their influence, these frameworks share a blind spot: they treat venues as passive constraints. In El Farol, the bar is just a threshold. In the minority game, the resource is an abstract binary choice. In KPR, restaurants are capacity-one slots with no agency whatsoever.

Real venues do not work like this. Bars adjust prices. They run happy hour specials. They impose cover charges on busy nights and use reservation systems to manage crowds. Restaurants implement dynamic pricing, manage waitlists, coordinate through platforms like OpenTable. Cloud providers auto-scale resources and adjust spot pricing based on demand. Each faces sustainability constraints-minimum revenue to stay in business, occupancy targets, operational costs-completely ignored in the classical models.

This matters because it misses half the game. In Arthur's model, coordination failure happens when customers can not predict each other. But in reality, the institution is also trying to predict and influence customer behavior. The bar might raise prices when it expects crowds or offer discounts to boost slow nights. These interventions change the coordination problem itself.

So the main questions are: what happens when we give the venue agency; when it has objectives, constraints, and learning capabilities just like the customers; when both sides adapt strategically to each other. The next sections build this model systematically.

\section{The Bar as Strategic Player}

In this section, we discuss the bar's strategy space, its corresponding objectives and payoffs, the formulation of the two-sided game, as also the mathematical model that describes it.

\subsection{Bar's Strategy Space}

We model the bar as an active decision-maker with control over several strategic variables. The primary control is pricing: the bar sets an effective price $p_t$ at time $t$, potentially varying it across periods to influence demand. This price may be decomposed as:

\begin{equation}
	p_t = p_0 - d_t
\end{equation}

where $p_0$ is a baseline price and $d_t \in [0, d_{\max}]$ is a discount level chosen by the bar.

Beyond pricing, the bar may control other strategic variables including capacity management (adjusting available seating or service levels), information disclosure (i.e., revealing or concealing attendance statistics to influence agent beliefs), and promotional policies (i.e., targeted discounts to specific customer segments). For analytical tractability, we focus primarily on the pricing mechanism as the bar's primary strategic tool.

\begin{figure}[t!]
	\centering
	\includegraphics[width=0.9\textwidth]{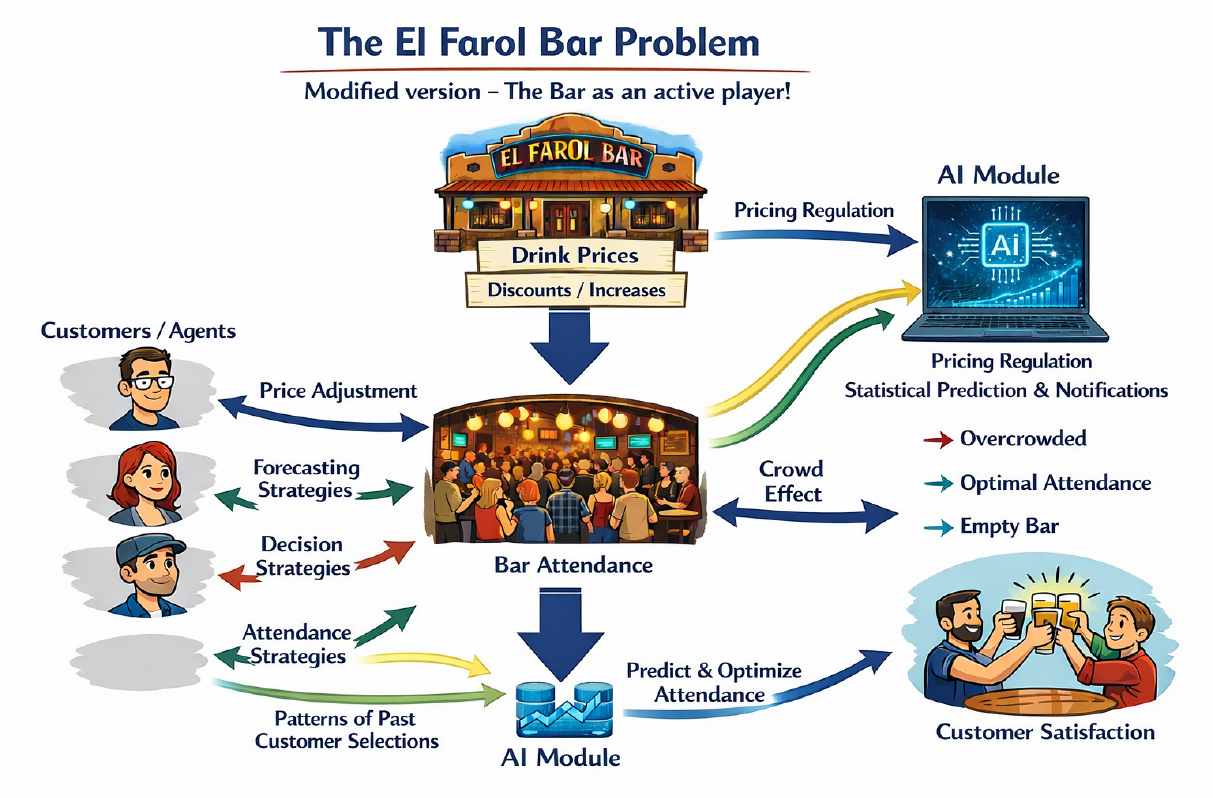}
	\caption{Modified El Farol Bar Problem with an AI-driven pricing mechanism. 
		The bar acts as an active player, adjusting drink prices based on predictions 
		from an AI module trained on past customer behavior, in order to optimize 
		attendance and customer satisfaction.}
	\label{fig:el_farol_ai}
\end{figure}

As shown in Figure~\ref{fig:el_farol_ai}, the AI module creates a feedback loop 
between customer behavior and pricing decisions, enabling adaptive regulation 
of attendance. The diagram presents a modified version of the El Farol Bar Problem, in which the bar is no longer a passive environment but an active decision-making agent. The system consists of three main components: customers (agents), the bar, and an AI module that mediates pricing decisions.

On the left side of the diagram, multiple customers/agents are depicted. Each agent follows different behavioral mechanisms, represented by arrows labeled forecasting strategies, decision strategies, and attendance strategies. These arrows indicate that agents form expectations about future crowd levels, make decisions based on those expectations, and ultimately decide whether to attend the bar. Additionally, a price adjustment interaction is shown, highlighting that customers respond to changes in drink prices set by the bar.

At the center of the diagram lies the bar, which is explicitly modeled as an active player. The bar controls drink prices through discounts and increases, represented by a downward arrow influencing bar attendance. This establishes the bar as a regulator of demand rather than a passive recipient of customer choices.

Below the bar, an AI module is introduced as a key extension of the classical problem. This module receives input in the form of patterns of past customer selections, capturing historical attendance behavior and individual decision patterns. Using this data, the AI performs statistical prediction and optimization, producing forecasts of expected attendance levels. Based on these predictions, the AI generates recommendations for pricing adjustments.

On the right side, a second representation of the AI module emphasizes its operational role in pricing regulation, where it sends signals back to the bar. This creates a feedback loop: the AI analyzes past behavior → predicts attendance → adjusts pricing → influences future customer decisions.

The outcome of this dynamic system is illustrated through the concept of crowd effects, which categorize attendance into three regimes: overcrowded, optimal attendance, and empty bar. These states directly impact customer satisfaction, shown at the bottom right of the diagram, where higher satisfaction is associated with balanced, optimal attendance levels. 

\noindent The diagram captures an adaptive system where:
\begin{itemize}
	\item Customers adapt their strategies based on expectations and prices,

	\item The bar actively adjusts prices to regulate demand,

	\item The AI module learns from historical patterns to optimize attendance.
\end{itemize}

This transforms the original El Farol Bar Problem into a data-driven, feedback-controlled system, where equilibrium emerges from the interaction between boundedly rational agents and an adaptive pricing mechanism.

\subsection{Bar's Objectives and Payoffs}

\noindent The bar pursues multiple, sometimes conflicting objectives. Let $K_t$ denote attendance at time $t$ and $p_t$ the effective price. Gross revenue is:

\begin{equation}
	R_t = p_t K_t
\end{equation}

The bar incurs costs $C_B(K_t)$ that may be increasing in attendance due to service requirements, or non-monotonic if very low attendance creates inefficiencies. Net profit is:

\begin{equation}
	\Pi_t = R_t - C_B(K_t) = p_t K_t - C_B(K_t)
\end{equation}

\noindent The bar faces several constraints:

\begin{itemize}
	\item	\textbf{Capacity constraint}: Attendance must not exceed maximum capacity:
			\begin{equation}
				K_t \leq K_{\max}
			\end{equation}
	\item	\textbf{Minimum occupancy}: To maintain atmosphere and operational efficiency:
			\begin{equation}
				K_t \geq K_{\min}
			\end{equation}
	\item	\textbf{Sustainability constraint}: Revenue must exceed a minimum threshold to avoid bankruptcy:
			\begin{equation}
				\Pi_t \geq \Pi_{\min}
			\end{equation}
\end{itemize}

\noindent The bar's objective is to maximize long-run expected profit:

\begin{equation}
	\max_{\{d_t\}} \mathbb{E}\left[\sum_{t=0}^{\infty} \beta^t \Pi_t \right]
\end{equation}

subject to the constraints above, where $\beta$ is a discount factor. This formulation balances short-term revenue against long-term sustainability.

\subsection{Two-sided Game Formulation}

The interaction between agents and the bar constitutes a two-sided dynamic game. On one side, a population of $n$ agents indexed by $i \in \{1,\dots,n\}$ repeatedly decides whether to attend. Let $a_i^t \in \{0,1\}$ denote agent $i$'s attendance decision at time $t$, with total attendance $K_t = \sum_i a_i^t$.

Each agent derives utility from attending depending on congestion and price:

\begin{equation}
	u_i^t = a_i^t \cdot [U(K_t) - p_t]
\end{equation}

where $U(K_t)$ is a satisfaction function capturing congestion effects, typically concave or even decreasing when $K_t$ exceeds comfortable levels.

On the other side, the bar chooses a discount policy $\{d_t\}$ to maximize its objective function subject to constraints. Crucially, both sides operate under incomplete information and bounded rationality. Agents do not observe the full attendance set or the bar's cost function. The bar does not perfectly know agent preferences or learning algorithms. Both must learn from historical data and adapt strategies over time.  This creates a dynamic game where:

\begin{itemize}
	\item	\textbf{Lower level}: Agents choose attendance strategies $\{\sigma_i\}$ given the bar's pricing policy $\pi = \{d_t\}$ and their beliefs about other agents' behavior.
	\item	\textbf{Upper level}: The bar chooses pricing policy $\pi$ anticipating how agents will respond through their learning dynamics.
\end{itemize}

\noindent The equilibrium of this game is not a static Nash equilibrium but rather a learning-consistent dynamic equilibrium where both agent strategies and bar policies co-evolve.

\subsection{Mathematical model}

We formalize the model structure as follows. Time is discrete, $t = 1,2,\ldots$ Agents observe only partial information: when agent $i$ attends at time $t$, they observe a subset $S_i^t \subseteq \{1,\ldots,n\} \setminus \{i\}$ of other attendees, with $|S_i^t| = K_t - 1$. Agents who do not attend observe nothing.

Each agent maintains a belief state $b_i^t$ representing their estimate of attendance probabilities for other agents, formed from historical observations $\{S_i^\tau\}_{\tau < t}$. Beliefs are updated via learning mechanisms described in Section~\ref{sec:ai}.

Agent $i$'s strategy is a mapping from their belief state and current price to an attendance probability:

\begin{equation}
	\sigma_i^t: (b_i^t, p_t) \to [0,1]
\end{equation}

\noindent The bar's state includes aggregate attendance history $\{K_\tau\}_{\tau < t}$ and revenue history $\{R_\tau\}_{\tau < t}$. The bar's policy maps this history to a discount level:

\begin{equation}
	\pi^t: (\{K_\tau, R_\tau\}_{\tau < t}) \to [0, d_{\max}]
\end{equation}

\noindent The system evolves through the following sequence each period:
\begin{enumerate}
	\item	Bar observes state and chooses discount $d_t$ via policy $\pi^t$, determining price $p_t = p_0 - d_t$
	\item	Agents observe price $p_t$ and form attendance decisions based on beliefs $b_i^t$ and strategies $\sigma_i^t$
	\item	Attendance realizes as $K_t = \sum_i a_i^t$
	\item	Agents receive payoffs $u_i^t = a_i^t \cdot [U(K_t) - p_t]$
	\item	Bar receives payoff $\Pi_t = p_t K_t - C_B(K_t)$
	\item	Attending agents observe subsets $S_i^t$ and update beliefs
	\item	Bar observes $K_t$, $R_t$ and updates policy
\end{enumerate}

\noindent This formulation captures the essential features: partial observability, sequential decision-making, learning dynamics on both sides, and sustainability constraints. The next section details how AI enables both agents and the bar to navigate this complex environment.

\section{AI-augmented Bar Intelligence} \label{sec:ai}

Next, we present the basis behind the utilization of AI-augmented decision support deployed for the bar, the capabilities provided to the bar by the deployment of Artificial-Intelligence, the learning provided to the customer, as also the co-evolutionary dynamics between the bar and the customers.

\subsection{Rationale for AI-Augmented Bar Systems}

The bar's optimization problem is genuinely hard, requiring to forecast demand from a population of customers who are themselves learning and adapting. Agent behavior is not stationary-their strategies evolve in response to past experiences, other customers' actions, and the bar's own pricing decisions. Hence, the bar can not just estimate a demand curve and optimize against it, while the curve itself keeps shifting as people learn.

Moreover, the bar only observes aggregate outcomes, and sees total attendance each week, but not individual customer types, preferences, or learning algorithms, without directly measuring the satisfaction function $U(\cdot)$ or knowing how price-sensitive different customers are. Everything has to be inferred from the correlation between prices and realized attendance over time.

Additionally, the bar is juggling multiple objectives under constraints-maximizing revenue while respecting capacity limits, maintaining minimum occupancy, avoiding bankruptcy. Classical optimization assumes stationary environments and known demand functions. Neither applies here.

AI provides a way through. Machine learning can forecast non-stationary demand and adapt as customer behavior changes. Reinforcement learning can discover effective pricing policies through trial and error without requiring a perfect model. Bayesian methods can quantify uncertainty about customer preferences. These are not just computational tools-they're fundamentally suited to the problem's structure: learning from limited data in a dynamic, uncertain environment.

\subsection{Bar-side AI Capabilities}
Next, we analyze to a further extent the forecasting regarding the capabilities provided from AI to the bar, the dynamic pricing and capacity optimization features, as also the learning characteristics in customer response patterns.

\subsubsection{Demand Forecasting}

The bar's first AI capability is demand forecasting: predicting attendance $K_t$ given price $p_t$ and historical data. This is a supervised learning problem where the bar observes pairs $(p_\tau, K_\tau)$ for $\tau < t$ and learns the conditional distribution $P(K_t | p_t, \text{history})$.

A simple approach uses a regression model:

\begin{equation}
	\hat{K}_t = f(p_t, K_{t-1}, \ldots, K_{t-M})
\end{equation}

where $f(\cdot)$ might be a linear model, random forest, or neural network trained on historical data. More sophisticated approaches model the non-stationarity explicitly, using time-varying coefficients or online learning methods that discount older observations.

The bar may also forecast uncertainty, not just expected attendance. A Bayesian approach maintains a posterior distribution over demand parameters, allowing the bar to quantify prediction confidence and make risk-aware decisions. For instance, Gaussian process regression provides both a mean prediction and a variance estimate, enabling the bar to balance exploration (testing new prices to learn demand) and exploitation (using known prices to maximize revenue).

\subsubsection{Dynamic Pricing and Capacity Optimization}

Given demand forecasts, the bar must choose prices to optimize its objective. This is a dynamic programming problem. The state at time $t$ includes attendance history, revenue history, and potentially the bar's belief state about agent preferences. The action is the discount $d_t$. The reward is the profit $\Pi_t$ subject to constraints.

A reinforcement learning approach frames this as a Markov decision process (MDP). The bar learns a policy $\pi^*: \text{state} \to \text{action}$ that maximizes expected cumulative discounted reward:

\begin{equation}
	\pi^* = \arg\max_\pi \mathbb{E}_{\pi}\left[\sum_{t=0}^{\infty} \beta^t \Pi_t \right]
\end{equation}

Methods like Q-learning, policy gradient algorithms, or actor-critic approaches can discover effective policies through trial and error. The bar experiments with different prices, observes outcomes, and adjusts its strategy to improve long-run performance.

Constraints (capacity, minimum occupancy, sustainability) can be incorporated through constraint violations as negative rewards or via constrained RL algorithms that explicitly enforce feasibility. For instance, if $K_t > K_{\max}$, the bar might incur a large penalty, incentivizing the policy to avoid overcrowding.

\subsubsection{Learning Customer Response Patterns}

Beyond forecasting aggregate demand, the bar can learn how agents respond to pricing changes. This involves identifying price elasticity: how sensitive is attendance to price variations? Do agents respond immediately or with a lag? Are there threshold effects where small price changes trigger large attendance shifts?

The bar can estimate a response function:

\begin{equation}
	\Delta K_t = g(\Delta p_t, \text{history})
\end{equation}

capturing how changes in price affect changes in attendance. Time-series methods like vector autoregressions (VAR) or dynamic linear models can capture lagged effects and feedback dynamics.

Additionally, the bar may learn agent heterogeneity. If the bar can segment customers (e.g., regulars vs. newcomers, price-sensitive vs. experience-focused), it can tailor pricing strategies. This might involve clustering observed attendance patterns or using mixture models to infer latent customer types.

\subsection{Customer-side Learning}

Agents also employ AI to navigate uncertainty. Each agent observes only partial information-subsets of past attendees-and must infer attendance patterns and predict future congestion.

\begin{itemize}
	\item	\textbf{Belief formation under partial observability}: Agents use Bayesian inference to form beliefs about other agents' attendance probabilities. Given observed sets $\{S_i^\tau\}_{\tau < t}$, agent $i$ estimates the probability that agent $j$ attends:
			\begin{equation}
				\hat{p}_{ij}^t = P(a_j^t = 1 | \{S_i^\tau\}_{\tau < t})
			\end{equation}
	This can be computed via empirical frequency (fraction of past observations where $j$ appeared) or more sophisticated Bayesian updating with priors on attendance probabilities. To this point, it is notable to refer that agent $i$ has observations only for the times went to the bar, that may require further investigation in selection bias as such cases may concern times where the bar has less attendance.
	\item	\textbf{Strategy adaptation via reinforcement learning}: Agents may use RL to learn attendance strategies directly. Each agent treats the environment as a Markov game, with states encoding their beliefs and the current price, actions being attend/not attend, and rewards being the payoff $u_i^t$. Algorithms like Q-learning or policy gradients allow agents to improve their strategies over time without explicitly computing Nash equilibria \cite{franke2003reinforcement}.
	\item	\textbf{No-regret learning}: Alternatively, agents might employ no-regret algorithms like multiplicative weights or Follow-the-Regularized-Leader (FTRL). These methods guarantee that the agent's time-averaged payoff approaches the best fixed strategy in hindsight. Under certain conditions, if all agents use no-regret learning, the empirical distribution of play converges to a coarse correlated equilibrium \cite{hart2000simple}.
\end{itemize}

\subsection{Co-evolutionary Dynamics}

When both customers and the bar learn simultaneously, you get feedback loops. Customer strategies evolve in response to the bar's pricing. The bar's pricing evolves in response to aggregate customer behavior. Neither reaches a fixed point; both continually adapt.

Consider a concrete example. Suppose the bar offers a large discount to boost attendance. More customers show up, which increases revenue-good. But it also creates overcrowding, which tanks customer satisfaction. Next week, customers who had a bad experience reduce their attendance probability. Lower attendance means lower revenue, which prompts the bar to increase discounts again. The cycle continues.

This can produce different dynamics depending on parameters. With slow learning rates and weak coupling between the two sides, the system might gradually settle near an approximate equilibrium where the bar's policy stabilizes and customer patterns become predictable. With faster learning or stronger interactions, you might get persistent oscillations or even chaos. The bar keeps chasing a moving target that it's simultaneously creating.

Whether this co-evolution leads to good outcomes is an open question. Under certain conditions-think slow, cautious learning on both sides-the system approaches something reasonable. In other regimes, it stays turbulent indefinitely.

Analyzing these dynamics requires tools from dynamical systems and evolutionary game theory. Mean-field approximations can collapse the customer population into a representative agent or a distribution over types. Then you analyze how the bar's policy responds to that aggregate and derive stability conditions. It's messy, but tractable in simplified cases.

\section{Equilibrium Analysis}

Next, we present the concept regarding the equilibrium with respect to the strategic bar, the properties among classical and AI-augmented equilibria, and provide an analysis regarding the welfare and the efficiency provided by the proposed approach.

\subsection{Equilibrium Concepts with Strategic Bar}

Classical Nash equilibrium assumes everyone simultaneously picks their best strategy given everyone else's fixed strategy. This does not fit our model for several reasons. The game repeats over time with strategies evolving. Information is incomplete and asymmetric-customers do not know each other's types or learning algorithms. And the bar's role creates a hierarchy: it sets prices anticipating customer responses, while customers choose attendance given those prices and their beliefs about others.

Instead we use learning-based equilibrium concepts. An approximate Bayesian Nash equilibrium happens when agents best-respond to their beliefs about others, and those beliefs are roughly correct. Formally, agent $i$'s strategy $\sigma_i$ is an $\varepsilon$-best response if:

\begin{equation}
	\mathbb{E}_{a_{-i} \sim b_i}[u_i(a_i, a_{-i})] \geq \max_{a_i'} \mathbb{E}_{a_{-i} \sim b_i}[u_i(a_i', a_{-i})] - \varepsilon
\end{equation}

where $b_i$ is agent $i$'s belief about what others will do. An approximate BNE requires all agents play $\varepsilon$-best responses and beliefs roughly match reality.

A weaker but useful concept is coarse correlated equilibrium (CCE). It just requires that no agent can improve by unilaterally switching to some fixed alternative strategy, where expectations are over the empirical distribution of past play. If everyone uses no-regret learning, the time-averaged play converges to the CCE set \cite{hart2000simple}-a nice result that connects learning algorithms to equilibrium outcomes.

With the bar as a strategic player, we also consider Stackelberg equilibrium. The bar acts as a leader, committing to a pricing policy. Customers respond optimally to that policy. The bar chooses its policy to maximize profit knowing how customers will react. This makes sense when the bar can credibly commit to prices (e.g., by posting them publicly), while customers retain short-term flexibility.

\subsection{AI-augmented vs Classical Equilibria}

In classical analysis, agents are assumed to compute optimal strategies with perfect knowledge of the game. In our AI-augmented model, agents instead learn through experience and approximate optimal behavior. This matters in several ways.

\begin{itemize}
	\item	\textbf{Equilibrium becomes an emergent property}. We do not assume agents start at equilibrium. Instead, equilibrium-if it arises-emerges from learning dynamics. The analytical focus shifts from solving for equilibrium to characterizing whether and how learning converges.
	\item	\textbf{Bounded rationality means approximate solutions}. AI-based learning typically produces approximately optimal strategies, not exact ones. Agents might use simple heuristics, limited memory, or coarse state representations. The resulting equilibria are approximate-$\varepsilon$-equilibria where $\varepsilon$ reflects computational limitations.
	\item	\textbf{The environment is non-stationary}. Classical equilibrium assumes things doo not change. But here, the bar's policy changes as it learns, and customer strategies change in response. This can prevent exact convergence. Instead of a fixed point, you get persistent fluctuations or limit cycles. The relevant concept becomes a statistical equilibrium-a distribution over states the system visits recurrently.
	\item	\textbf{Beliefs do not necessarily converge to rational expectations}. Classical Bayesian Nash equilibrium assumes common priors and Bayesian updating. In practice, agents might have different priors, use biased learning algorithms, or form beliefs from limited data. Neural networks and kernel methods introduce their own inductive biases. Equilibrium beliefs can systematically differ from what rational expectations would predict.
\end{itemize}

\subsection{Welfare and Efficiency Analysis}

In order to  compare the outcomes provided from a strategic bar or a passive one, we deploy the following welfare metrics:

\noindent \textbf{Agent surplus}: total utility across all customers:

\begin{equation}
	W_{\text{agents}}^t = \sum_{i=1}^{n} u_i^t = \sum_{i: a_i^t = 1} [U(K_t) - p_t]
\end{equation}

\noindent \textbf{Bar profit}: $\Pi_t = p_t K_t - C_B(K_t)$.

\noindent \textbf{Social welfare}: sum of agent surplus and bar profit:

\begin{equation}
	W_{\text{total}}^t = W_{\text{agents}}^t + \Pi_t = \sum_{i: a_i^t = 1} U(K_t) - C_B(K_t)
\end{equation}

Notice prices cancel in social welfare-they're just transfers from customers to the bar. Social efficiency depends only on matching attendance to the satisfaction and cost functions.

When the bar is passive (fixed price and capacity), attendance is determined purely by customer coordination. If customers fail to coordinate well, attendance might frequently overshoot or undershoot the efficient level $K^*$ that maximizes total value. Strategic pricing can improve efficiency by steering attendance toward $K^*$. If crowds tend to overshoot, raise prices to reduce demand. If they undershoot, offer discounts.

But strategic pricing also creates losses. The bar only cares about its own profit, not customer surplus. With market power, it might set prices above marginal cost, reducing attendance below the socially efficient level to extract more revenue from customers who still show up. Standard monopoly distortion.

The net welfare effect depends on the balance: does improved coordination outweigh monopoly distortion? In low-variance settings where customers coordinate reasonably well already, giving the bar pricing power might just reduce surplus. In high-variance settings with frequent miscoordination, strategic pricing could improve things by stabilizing attendance.

There's also a dynamic dimension. If the bar faces bankruptcy constraints, strategic pricing might be necessary for survival. Even if a passive bar with low prices would generate positive social surplus in the short run, it might fail financially. Strategic pricing sustains the bar, preserving long-run access to the resource. Short-run monopoly distortions become acceptable if they enable long-run availability.

\section{Discussion and Future Directions}

In conclusion, we provide a discussion on the applications of the proposed approach and present the limitations observed throughout our analysis, setting also the corresponding open questions that guide our future research directions.

\subsection{Applications}

The framework described in this work applies to numerous real-world domains where service providers face congestion and must manage demand dynamically.

\begin{itemize}
	\item	\textbf{Dynamic pricing in hospitality}: Hotels, restaurants, and entertainment venues increasingly use AI-driven revenue management systems that adjust prices based on predicted demand. Our model provides a game-theoretic foundation for understanding how customers adapt to these pricing strategies and how providers should optimize policies when facing learning customers.
	\item	\textbf{Cloud resource allocation}: Cloud computing platforms allocate resources (compute instances, bandwidth, storage) to users who compete for limited capacity. Providers use spot pricing that fluctuates with demand. Our framework models the interaction between users learning to bid strategically and platforms learning to set prices that balance utilization and revenue.
	\item	\textbf{Congestion management in transportation}: Ride-sharing platforms adjust surge pricing to balance supply and demand. Drivers and riders learn these patterns and adjust their participation decisions. Our model captures this co-evolutionary dynamic and can inform surge pricing algorithms that account for behavioral adaptation.
	\item	\textbf{Network resource allocation}: In communication networks, users compete for limited bandwidth. Network operators can implement congestion pricing or quality-of-service tiers. Our framework analyzes how users' learning about network conditions interacts with operators' adaptive pricing and capacity management.
\end{itemize}

\subsection{Limitations and Open Questions}

The approach proposed in this work is simplified in ways that matter for real applications.

\begin{itemize}
	\item	\textbf{Stylized setting}: We focus on a single bar with homogeneous customers and one pricing tool. Real markets have heterogeneous customers with different valuations, multiple competing venues, and richer strategy spaces (reservation systems, capacity expansion, loyalty programs). Extending the model to these settings would help, but also complicates analysis significantly.
	\item	\textbf{Convergence is assumed, not proven}: We discuss learning dynamics conceptually, but rigorous convergence analysis is sparse. Under what conditions do customer learning and bar policy learning jointly converge to equilibrium? What learning rates, algorithm choices, and parameters ensure stability? This needs dynamical systems analysis and stochastic approximation theory-technical work we've mostly sidestepped.
	\item	\textbf{Learned vs optimal mechanisms}: We model the bar as learning a pricing policy through trial and error. An alternative is designing the optimal mechanism directly-solving the bar's problem analytically given assumptions about customer behavior. How close do learned policies come to theoretical optima? Can mechanism design principles inform algorithm design? We do not know yet.
	\item	\textbf{Fairness matters}: Strategic pricing can hurt price-sensitive customers or create unequal access. How do we incorporate fairness constraints into the bar's optimization? Can AI-based pricing balance efficiency, revenue, and equity? These are normative questions the model currently ignores.
	\item	\textbf{Strategic manipulation}: If customers understand the bar's learning algorithm, they might strategically manipulate observed data to induce favorable pricing. The bar might try to manipulate customer beliefs. Analyzing strategic information revelation and robust mechanism design in this context is important but unexplored here.
	\item	\textbf{No empirical validation}: The model makes testable predictions about how pricing and attendance co-evolve. But we haven not validated it against real data from restaurants, bars, or ride-sharing platforms. Empirical work could estimate key parameters and test whether the mechanisms we describe actually operate in practice.
\end{itemize}

In conclusion, re-conceptualizing the El Farol Bar problem with the bar as a strategic, learning-capable player bridges game theory, AI, and mechanism design in ways that matter for real systems. The framework provides both theoretical insights into multi-agent learning and practical tools for managing congestion and sustainability in markets where institutions adapt alongside customers.

\bibliographystyle{ieeetr}
\bibliography{EElFBG}

\end{document}